\documentclass[10pt,aps,pre,twocolumn,superscriptaddress]{revtex4-1}
\usepackage{amsmath,amssymb,amsthm,mathrsfs,amsopn}
\usepackage{graphicx}
\usepackage{subfigure}
\usepackage{booktabs}
\usepackage{array}
\usepackage{tcolorbox} 
\tcbset{colback=blue!5!white,colframe=blue!50!black, left=0mm,right=0mm,top=0mm,bottom=0mm, box align=center, halign=center} 
\usepackage{tikz} 
\usepackage{enumitem}
\usepackage{fancyhdr} 

\begin{document}

\title{Impact of network characteristics on network reconstruction}
\author{Gloria Cecchini}
\thanks{These two authors contributed equally}
\affiliation{CSDC, Department of Physics and Astronomy, University of Florence, Sesto Fiorentino, Florence, Italy}
\affiliation{Institute of Physics and Astronomy, University of Potsdam, Campus Golm, Karl-Liebknecht-Stra{\ss}e 24/25, 14476 Potsdam-Golm, Germany}
\author{Rok Cestnik}
\thanks{These two authors contributed equally}
\affiliation{Institute of Physics and Astronomy, University of Potsdam, Campus Golm, Karl-Liebknecht-Stra{\ss}e 24/25, 14476 Potsdam-Golm, Germany}
\affiliation{Faculty of Behavioural and Movement Sciences, Vrije Universiteit Amsterdam, Van der Boechorststraat 7, 1081 BT Amsterdam, Netherlands}
\author{Arkady Pikovsky}
\affiliation{Institute of Physics and Astronomy, University of Potsdam, Campus Golm, Karl-Liebknecht-Stra{\ss}e 24/25, 14476 Potsdam-Golm, Germany}
\affiliation{Department of Control Theory, Lobachevsky University of Nizhny Novgorod, Gagarin Av. 23, 603950, Nizhny Novgorod, Russia}

\date{\today}

\begin{abstract}
When a network is inferred from data, two types of errors can occur: false positive and false negative conclusions about the presence of links.
We focus on the influence of local network characteristics on the probability $\alpha$ - of \textit{type I} false positive conclusions, and on the probability $\beta$ - of \textit{type II} false 
negative conclusions, in the case of networks of coupled oscillators. We demonstrate that false conclusion probabilities are influenced by local connectivity measures such as the shortest path length and the detour degree, which can also be estimated from the inferred network when the true underlying network is not known a priory. These measures can then be used for quantification of the confidence level of link conclusions, and for improving the network reconstruction via advanced concepts of link thresholding. 
\end{abstract}

\maketitle

\section{Introduction}

Complex systems are of key interest in multiple scientific fields, ranging from medicine, physics, mathematics, engineering, economics etc. \cite{Barrat2008,Boccaletti2006,Cohen2010,PIK01}.
Many complex systems can be modeled, or represented as dynamical networks, where nodes are the dynamical elements and links represent the interactions between them. 
In this context, networks are widely used in studies of synchronization phenomena of coupled oscillators as well as in the analysis of chaotic behavior in complex dynamical systems \citep{Rok2017,Arkady,Li2014,Arkady2016}. 
A deep understanding of network characteristics allows controlling the network dynamics \citep{Bahadorian2019}, e.g., in case of optimizing vaccination strategies with the aim of controlling the spread of diseases \citep{Clusella2016}.
Very often one faces an \textit{inverse problem}: the underlying network is not known, and a reliable inferring of the network structure from the observation is crucial for understanding the system's operation \citep{Lehnertz2016,Casadiego2018,Timme2016,Casadiego2015,Kutz2016,Kutz20161,Leguia20191,Arkady2011,Leguia_2017,adam2019}.

When a network is to be inferred from observation data, typical analysis techniques provide measures of connectivity strength for each link.
Several methods have been suggested in the literature to reconstruct the network structure and decide whether these measures pass a certain threshold, thereby providing a mean to decide if the corresponding links are considered as present or not \citep{Arkady2018,Asllani2018, Burioni2014, Shandilya2011, Leguia2019, Banerjee2019, Panaggio2019}.

If a non-existing link is erroneously detected, it is called a false 
positive link and is referred to as a \textit{type I} error.
Likewise, an existing link that remains undetected 
is called a false negative link and is referred to as a \textit{type II} error.
The probability of detecting a false positive link is usually denoted 
by $\alpha$, while $\beta$ denotes the probability that an existing 
link remains undetected. Of course, the goal of a reliable reconstruction 
is to minimize both these probabilities simultaneously.

In \citep{GloriaJNM2018, GloriaPRE2018,gloria2020}, the analysis of the errors of both types was focused on the 
influence of false positive and false negative conclusions about links 
on the reconstructed network  characteristics. 
It was demonstrated, that within the same network topology, the values for $\alpha$ and $\beta$, leading to the least biased network
characterisation, change depending on the 
network property of interest. 
In this paper, the analysis is reversed - the study focuses on the influence 
of network characteristics on the probabilities 
of \textit{type I} and \textit{type II} errors.

Below, we first assume the knowledge of the true underlying network. 
In Section~\ref{secDepFPNonChar} we perform a simulation study 
to show the dependence of the probability of false positive and false 
negative links on their shortest path length and their detour degree (defined 
later in section~\ref{sec:bin_networks}). 
In Section~\ref{secLast}, 
these results are applied to a scenario where the underlying network is 
unknown {\it a priori}, so we evaluate the shortest
path length and of the detour degree from the reconstruction
to improve the  quality of the latter, 
i.e. to decrease the number of falsely concluded links.

\section{Networks and Methods}
\label{chapterNetworkTheory}
In this section we give necessary network definitions.
A network is defined as a set of nodes with links between them \citep{NewmanBook}. 
In graph theory, a branch of mathematics that studies networks, a different notation is used: networks are called graphs, and nodes and links are called vertices and edges, respectively.
Below, the notations from network theory and graph theory are used synonymously.

In this paper, Erd{\H o}s-R{\'e}nyi networks are used for the simulation study.
Erd{\H o}s-R{\'e}nyi networks are random networks in which the set of nodes is fixed, and each pair of nodes is connected with independent probability $p$.
The probability mass function of the node degree distribution of an Erd{\H o}s-R{\'e}nyi network is a binomial distribution
\begin{equation}
\mathbb{P}(d_v=k)=\binom{n-1}{k}p^k(1-p)^{n-1-k}\;,
\label{massFunErdos}
\end{equation}
where $n$ is the number of nodes in the network.

\subsection{Binary networks}
\label{sec:bin_networks}

The adjacency matrix $\mathcal{A}$ of a binary network with $n$ nodes is an $n\times n$ matrix with elements
\begin{equation}
\mathcal{A}_{ij}=
\begin{cases}
1 \text{ if there is link from node $i$ to node $j$,}\\
0 \text{ otherwise.}
\end{cases}
\end{equation}

Networks can be directed or undirected.
In an undirected network, connection from $i$ to $j$ implies the connection from $j$ to $i$.
Note that this implies that the adjacency matrix is symmetric.
In a directed network, this symmetry is broken, therefore if a path from $i$ to $j$ exists, a path from $j$ to $i$ does not necessarily exist. 
We will consider directed networks and hence non-symmetric adjacency matrices.

For two randomly selected nodes $i, j$ in a network of $n$ nodes, the shortest path length (SPL) $\ell_{ij}$ measures the number of links separating them if the shortest path is taken. 
For connected nodes $i, j$, when the oriented edge $i \rightarrow j$ exists, the SPL is $\ell_{ij}=1$. For directed networks generally $\ell_{ij}\neq \ell_{ji}$. 

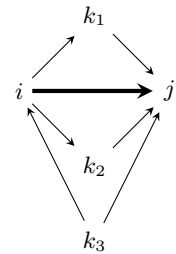
\begin{figure}[!thb]
\centering
\begin{tikzpicture}[[->,>=stealth,shorten >=1pt,auto]
\node (j) at (0:1) {$j$};
\node (k1) at (90:1) {$k_1$} edge [->] (j);
\node (i) at (180:1) {$i$} edge [->, ultra thick] (j)
edge [->] (k1);
\node (k2) at (270:1) {$k_2$} edge [->] (j);
\path[->] (i) edge (k2);
\node (k3) at (270:2) {$k_3$} edge [->] (j);
\path[<-] (i) edge (k3);
\end{tikzpicture}
\caption{Example of DD $\Delta_{ij}=2$.}
\label{figExDetour}
\end{figure}

Inspired by the idea of a local clustering coefficient \citep{NewmanBook}, 
a novel network characteristic, 
which we refer to as the \textit{detour degree} (DD) $\Delta_{ij}$, 
is defined here.
Detour degree is a pairwise measure that quantifies detours between a 
pair of nodes. 
Namely, for every oriented node pair $i\rightarrow j$, the detour degree is the number of oriented paths of length 2 from $i$ to $j$.
For example, in the case shown in Fig.~\ref{figExDetour}, the DD is $\Delta_{ij}=2$, corresponding to two directed paths of length 2 from $i$ to $j$ through $k_1$ and $k_2$.
Since the edge between $i$ and $k_3$ is oriented towards $i$, a path from $i$ to $j$ through $k_3$ does not exist. 
Similarly to the SPL, the DD is non-symmetric for directed networks. Notice also some connection between the SPL and the DD: if $\ell_{ij}\geq 3$, then $\Delta_{ij}=0$.

\begin{figure*}[!ht] 
\centering
\includegraphics[width=\textwidth]{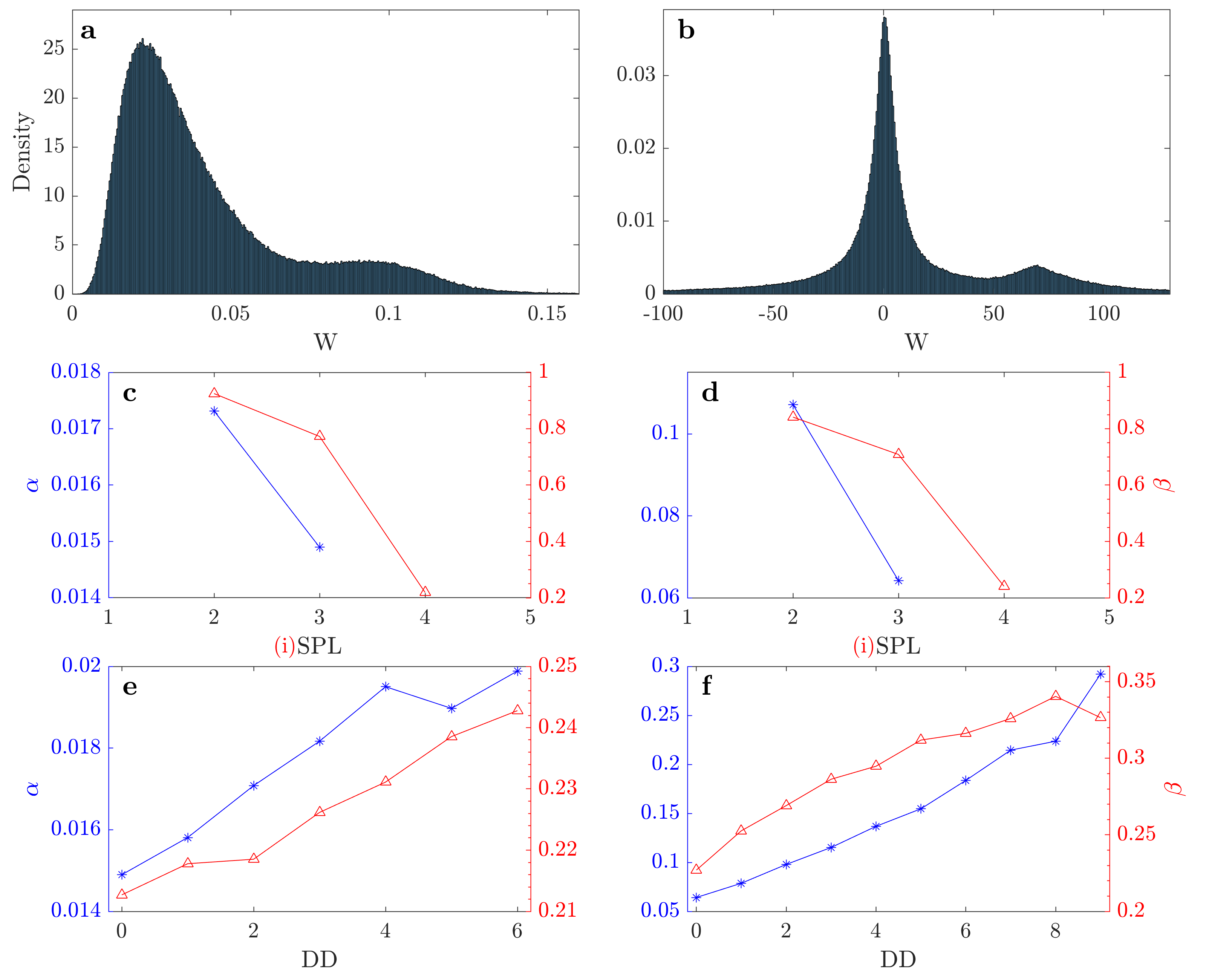}
\caption{Inferred coupling strengths, relationship of $\alpha$ and $\beta$ as function of the SPL and DD using two inference techniques: $G_1$ (panels a,c,e) and $G_2$ (panels b,d,f). Panels (a-b): Histograms of the inferred coupling strengths. Probabilities $\alpha$ and $\beta$ as functions of the SPL (c-d), and as functions of the DD (e-f), for a specific value of the threshold ($0.08$ for $G_1$ and $45$ for $G_2$).}
\label{figHistAlphaBeta}
\end{figure*}

\subsection{Weighted networks}
\label{secwn}

Often it is useful to define a network where 
the links are not binary connections, but are instead described by continuous 
weights. The adjacency matrix elements of weighted networks are 
real numbers. 
Definitions provided in the previous section for the SPL and 
the DD in binary networks are 
here generalized for weighted networks.

We consider the direct path length from node $i$ to node $j$ to be the inverse of the corresponding adjacency matrix element $\mathcal{A}_{ij}$ \citep{Opsahl}, or in other words, the inverse of the link weight. Therefore, the SPL from node $i$ to node $j$ is the minimal sum of pairwise path lengths for all available paths between $i$ and $j$, i.e. 
\begin{equation}
\ell_{ij}=\min\left(\mathcal{A}_{i k_1}^{-1}+\cdots+\mathcal{A}_{k_n j}^{-1}\right),
\label{shortPathWeight}
\end{equation} 
where nodes $k_1$ through $k_n$ belong to all 
possible paths from $i$ to $j$. 
Note that for binary networks, this definition is coherent with the one in the previous section.
For a binary network,
an existing link corresponds to weight 1 and an absent link to weight 0, the latter would lead to an infinite contribution in the sum.
Therefore Eq.~(\ref{shortPathWeight}) reduces to the number of links separating $i$ and $j$ if the shortest path is taken. 
As a sidenote, one can draw a parallel here with circuit theory~\citep{circuit_theory}, with link weights representing directed conductances, making the shortest path correspond to the path of least resistance and the SPL quantify its effective resistance. 

The DD of an oriented node pair $i\rightarrow j$ measures the contribution of all the possible 2-step paths from $i$ to $j$.
In weighted networks, such a contribution must consider the weights of the edges.
Namely, the DD is scaled by the product of weights of the two edges that form the 2-step path
\begin{equation}
\Delta_{ij}= \sum_k \mathcal{A}_{ik}\mathcal{A}_{kj}.
\label{detourWeight}
\end{equation}
For binary networks, this definition is coherent with the definition in the previous section, since for $\mathcal{A}_{kh}\in\{0,1\}$ Eq.~(\ref{detourWeight}) reduces to the total number of paths of length 2 from node $i$ to node $j$.
In the circuit theory analogy~\citep{circuit_theory}, the DD roughly corresponds to the effective conductance of all paths of length 2 (that would be $\sum_k \frac{\mathcal{A}_{ik}A_{kj}}{\mathcal{A}_{ik}+\mathcal{A}_{kj}}$).
Note that, in both the binary and weighted case, Eq.~(\ref{detourWeight}) can be expressed elegantly in matrix form as $\Delta=\mathcal{A}^2$.

\subsection{Network inference examples}
\label{secRecMeth}

It is not a goal of this study to develop a novel network inference method;
rather we take methods from the previous literature and consider 
how they are affected
by the network properties.
We perform our studies with two network inference techniques. 
The first one takes continuous signals of all oscillators and assumes they follow the Kuramoto model dynamics~\citep{PIK01}: 
\begin{equation}
\dot{\phi}_k=\omega_k+\epsilon\sum_j T_{kj}\sin(\phi_j-\phi_k-\Theta_{jk})
\end{equation}
where $\epsilon$ is the coupling 
strength, $\phi_k$ the phases, $\omega_k$ the natural frequencies 
and $\Theta_{jk}$ phase shifts. 
It returns strictly positive values for interactions $\epsilon T_{kj}$. 
For details see Ref.~\citep{Arkady2018}.
A network inferred using this technique is indicated in this manuscript with $G_1$, and Fig.~\ref{figHistAlphaBeta}a shows an example of inferred coupling strengths.

The second technique is designed for pulse-coupled oscillators. It takes the
observed spike times and assumes that the interaction can be well represented with a network based on the Winfree phase equation~\citep{winfree}:
\begin{equation}
\dot{\phi}_k=\omega_k+\epsilon Z_k(\phi_k)\sum_j T_{kj} \delta(t-t_{j})
\end{equation}
where $Z_k(\phi)$ is the 
phase response curve and $t_{k}$ are the spike times of oscillator $k$. 
The technique returns real numbers (positive and negative) for interactions. 
For details see Ref.~\citep{Rok2017}.
A network, inferred using this technique, is indicated in this 
manuscript with $G_2$, and Fig.~\ref{figHistAlphaBeta}b shows an 
example of inferred coupling strengths.

\section{Dependence of false conclusions on network characteristics}
\label{secDepFPNonChar}

This section focuses on the dependence of false positive and false 
negative link conclusions on the network characteristics introduced
in Section~\ref{chapterNetworkTheory}. 
To this aim we simulate an ensemble of oscillatory networks, and infer their 
connectivity from limited observations of its time series. 
We consider two different inference techniques, both of which yield 
continuous values for link weights, see Sec.~\ref{secRecMeth}. 

We denote the true network's binary adjacency matrix with $T$ and the 
inferred weighted one with $W$. 
The aim is to reconstruct the original binary 
network $T$ from the inferred one $W$, i.e. determine on the 
basis of link weights $W_{ij}$ whether the links are present or not. 
This is typically done by thresholding the weights, 
i.e. if an inferred link weight passes a certain threshold, the link is 
assumed to be present.

The inferred coupling strengths $W_{ij}$ have a certain distribution. 
Consequently, depending on the chosen threshold value, 
different numbers of false positive and false negative conclusions occur. 
This is commonly represented with a receiver operating 
characteristic, commonly referred to as a ROC curve~\citep{roc_curve}.
In this manuscript the interest is focused on the influence of 
the probabilities of false conclusions on the
local network characteristics SPL and DD.

The simulation study is performed on Erd{\H o}s-R{\'e}nyi 
networks with $n = 100$ nodes and probability of connection $p=0.15$.
In particular, for $G_1$ the frequencies $\omega_k$ are 
uniformly distributed within the interval $(0.5,1.5)$, the phase 
shifts $\Theta_{jk}$ are uniformly distributed in the 
interval $(0,2\pi)$, the original coupling strength is set 
to $\epsilon=0.3$, and 500 data points are used to perform 
the network inference. 
For $G_2$, the frequencies $\omega_k$ are uniformly distributed within 
the interval ($1.0,2.0)$, the coupling strength is set to $\epsilon=0.5$, 
all oscillators are assigned the same phase response curve: 
$Z(\varphi) = -\sin(\varphi) \exp(3 \cos(\varphi-0.9\pi))/\exp(3)$, 
and all 
spikes that occur within 500 observed periods 
of the slowest oscillator are considered for network inference. 
For both $G_1$ and $G_2$, 100 simulations are made to have enough 
statistical data.

\subsection{False conclusions with respect to local network structures}
\label{secFPspl}

In this section we study how the inferred weights, and therefore false conclusions, depend on the local characteristics of the true network $T$, namely the shortest path length (SPL) and the detour degree (DD). 
Since $T$ is discrete so too are the SPL and DD. 
It is worth noting here that we consider that all possible links $i \rightarrow j$ can be falsely identified regardless whether they are present in $T$ or not. Their presence simply determines whether they are candidates for a false positive conclusion (not present in $T$), or a false negative one (present in $T$).

The probability of a false positive conclusion $\alpha$ is evaluated for subsets of links with the same SPL: $\ell = 2, 3,...$ ($\ell = 1$ means the corresponding link exists and therefore no false positive conclusion can be made), see Fig.~\ref{figHistAlphaBeta}c-d. 
In the case of false negative conclusions however, the true link is present and the shortest path length therefore equal to 1. Because of this, we consider the indirect shortest path length (iSPL), i.e. SPL when the direct link is not considered - for clarity we distinguish its notation as $\tilde{\ell}$. Note that if $T_{ij} = 0$ then $\tilde{\ell}_{ij} = \ell_{ij}$. The probability of a false negative conclusion $\beta$ is then evaluated on links with the same iSPL: $\tilde{\ell} = 2, 3,...$ ($\tilde{\ell}$ of a binary network can not be less than 2). 
What we observe is that false conclusions happen more often for links with shorter (i)SPL. 
This intuitively makes sense. 
The smaller the (indirect) distance between two nodes the more they influence each other via indirect coupling, which can disrupt the inference algorithms~\citep{Arkady2018, Rok2017} into misinterpreting the connectivity. This holds true for both $\alpha$ and $\beta$. We depict these dependencies in Fig.~\ref{figHistAlphaBeta}c-d.

We perform a similar analysis using the DD in place of the SPL (Fig.~\ref{figHistAlphaBeta}e-f). 
The probabilities of false conclusions $\alpha$ and $\beta$ are evaluated for subsets of links with the same DD. 
We find that both $\alpha$ and $\beta$ typically increase with the DD. 
This again makes intuitive sense for the same reason as with the SPL.
Namely, if the DD is low, the indirect interaction between the nodes is low regardless of whether the direct connection exists or not. 
This means that there are less interferences to be picked up by the inference algorithms. 
These dependencies are depicted in Fig.~\ref{figHistAlphaBeta}e-f.

Here we point out that the DD is effectively a measure of connectivity while SPL is a measure of detachment, i.e. they measure opposite things. 
In circuit theory analogy DD is a measure of effective conductance while SPL is a measure of effective resistance.

\section{When the true graph is unknown}
\label{secLast}
\subsection{Using network characteristics from the reconstruction}

As we have seen in  Section~\ref{secFPspl}, false conclusion probability increases with the measure of indirect distance between nodes, i.e. it increases with the (i)SPL and decreases with the DD. 
The study presented above will be now  reversed - suppose the true network $T$ is not known and we only have access to the inferred wights $W$. 
In this section we investigate the possibility of using local network information of the inferred graph $W$ to gain additional insight on the probability of link existence.

\begin{figure*}[!thb]
\includegraphics[width=0.85\textwidth]{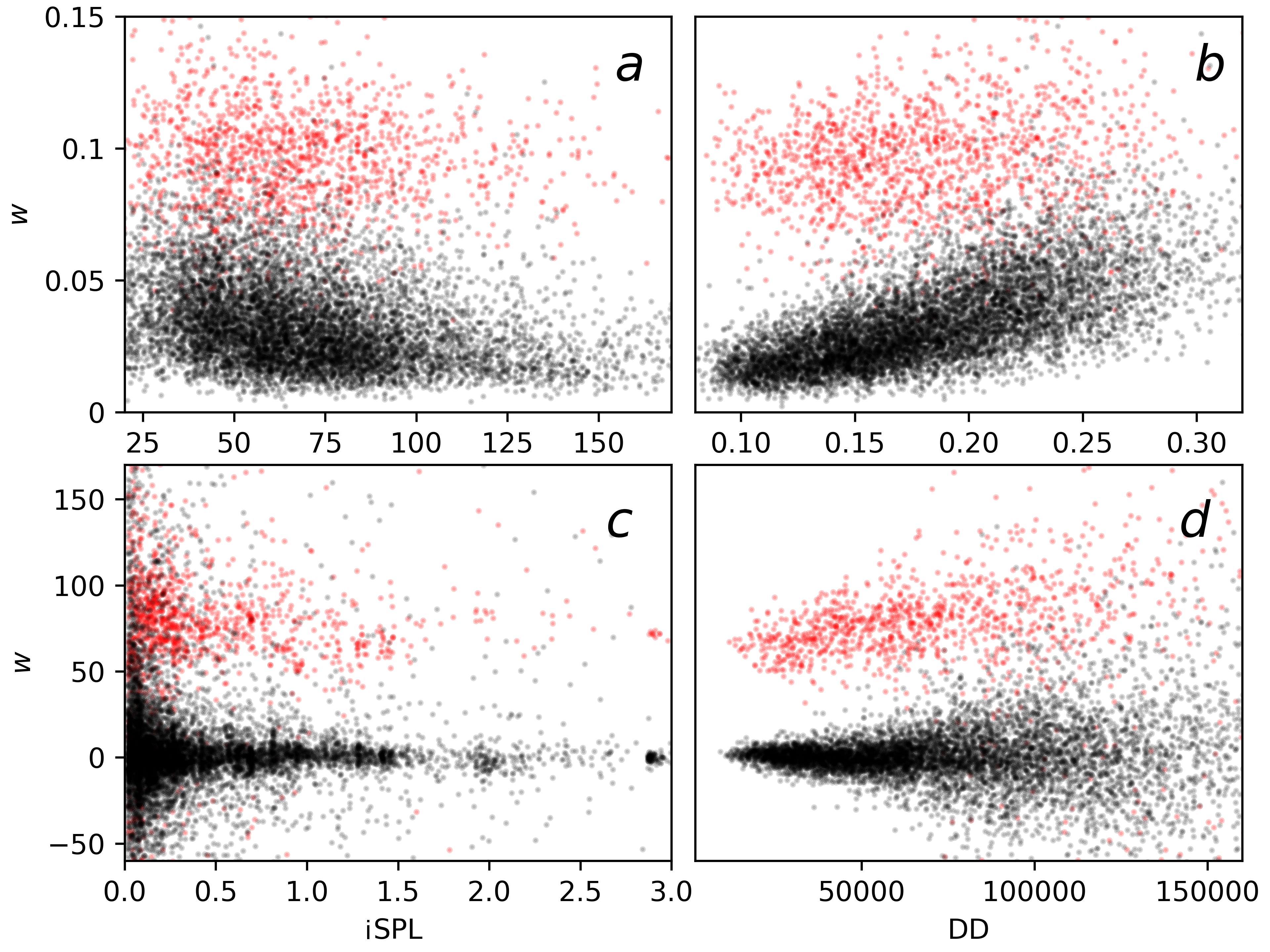}
\caption{Scatter plots of the inferred weights $w$ versus the iSPL (panels a,c) and versus the DD (panels b,d) for both inference methods, $G_1$ (panels a,b) and $G_2$ (panels c,d). Points corresponding to true links are depicted with red and false ones with black.}
\label{fig:scatters}
\end{figure*}

We can evaluate iSPL with Eq.~(\ref{shortPathWeight}), and DD with Eq.~(\ref{detourWeight}) on the inferred network $W$.
If any weights are negative we take their absolute value, the reasoning being that we are interested in the estimated interaction between nodes and negative weights represent a kind of interactions as well.  
Then we compare the relationships between the inferred link weight $W_{ij}$, the iSPL $\tilde{\ell}_{ij}$ and the DD $\Delta_{ij}$ - all obtained from $W$.

\begin{figure}[!thb]
\includegraphics[width=0.35\textwidth]{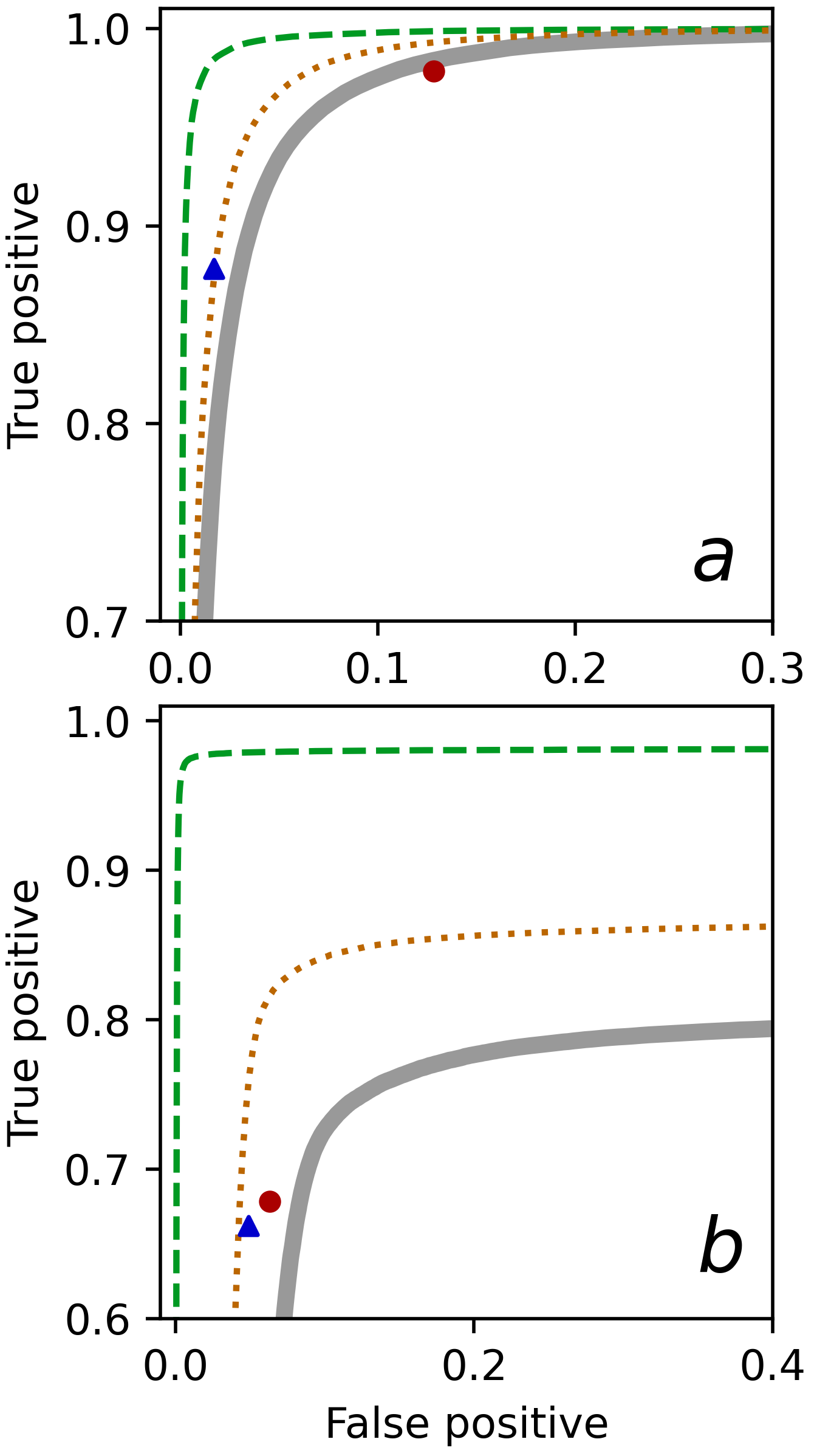}
\caption{ROC curves corresponding to: complete network reconstruction (thick gray line), 50\% of links with the lowest DD (dashed green line) and 50\% of links with the highest iSPL (dotted orange line). Best results
correspond to the upper left corner of the ROC plot.
The point corresponding to the {\it mountain-pass} 
thresholding is depicted with a blue triangle, 
and the one corresponding to the {\it SPL-relative} 
thresholding with a red circle. Both methods $G_1$ and $G_2$ 
are represented in panels a and b respectively. }
\label{fig:roc}
\end{figure}

In Fig.~\ref{fig:scatters}, we show scatter plots of weights $W_{ij}$ versus their corresponding links' iSPL (panels a-b), and versus DD (panels c-d), using the two network inference methods explained in Sec.~\ref{secRecMeth}.
We color the points differently for the ones that represent a true link, $T_{ij} = 1$ (red), and the ones that do not, $T_{ij} = 0$ (black). 
This reveals the qualitative dependence of weights on indirect measures of connectivity: iSPL and DD. 
The findings are reflective of those in Sec.~\ref{secFPspl}, namely, the probabilities of false conclusions decrease with iSPL and increase with DD. 
This means that these measures can be used to represent the level of confidence in detected links, i.e. links with low DD and high iSPL are more likely to be accurately reconstructed by thresholding.

We illustrate this with ROC curves evaluated on only a selected portion of links, according to their DD and iSPL. In particular, we consider the more confident half of links and compute false conclusions proportionally. These partial-consideration ROC curves are shown alongside the full-consideration curve as comparison, see Fig.~\ref{fig:roc}. The DD in particular seems to be a good indicator of confidence in a link conclusion.

\subsection{Alternative thresholding}
\label{secDiffTh}

The results presented in Sec. \ref{secFPspl} show the dependence of the inferred coupling strengths on two network characteristics - the indirect shortest path length (iSPL) and the detour degree (DD).
These results suggest that network reconstructions might benefit from different strategies of determining the existence of links. 
The na\"ive choice consists of selecting a threshold value, and considering all links with inferred coupling larger than the threshold as present, while the rest as not present. 
In this section, two advanced thresholding strategies are discussed.

The first possibility we discuss takes into account the relationship between the link's inferred coupling strength and its SPL. 
Specifically, one of many natural choices is to only consider links as present when their inverse coupling strength corresponds to their SPL. In other words, consider present all links for which the inferred SPL goes through the direct link. 
This choice can be graphically represented with a curved threshold, taking the $1/x$ curve in the plot Fig.~\ref{fig:scatters}a,c. 
We refer to this as the {\it SPL-relative} threshold.
Figure \ref{fig:roc} shows the ROC curve corresponding to the na\"ive choice for the threshold, and the circle red marker corresponds to the SPL-relative threshold.
While this does not seem to improve the reconstruction for $G_1$, it does significantly enhance the results for $G_2$. 
Further, we could consider combining SPL-relative threshold with the na\"ive threshold, by simply thresholding the remaining links. 
Namely, among the links whose strength corresponds to the reciprocal of the SPL, we perform simple thresholding. 
With this combined thresholding the reconstruction is marginally improved for $G_1$ as well, i.e. within a range of threshold values both $\alpha$ and $\beta$ are marginally reduced.

\begin{figure}[!t]
\centering
\begin{minipage}{0.435\textwidth}
\centering
\includegraphics[width=\textwidth]{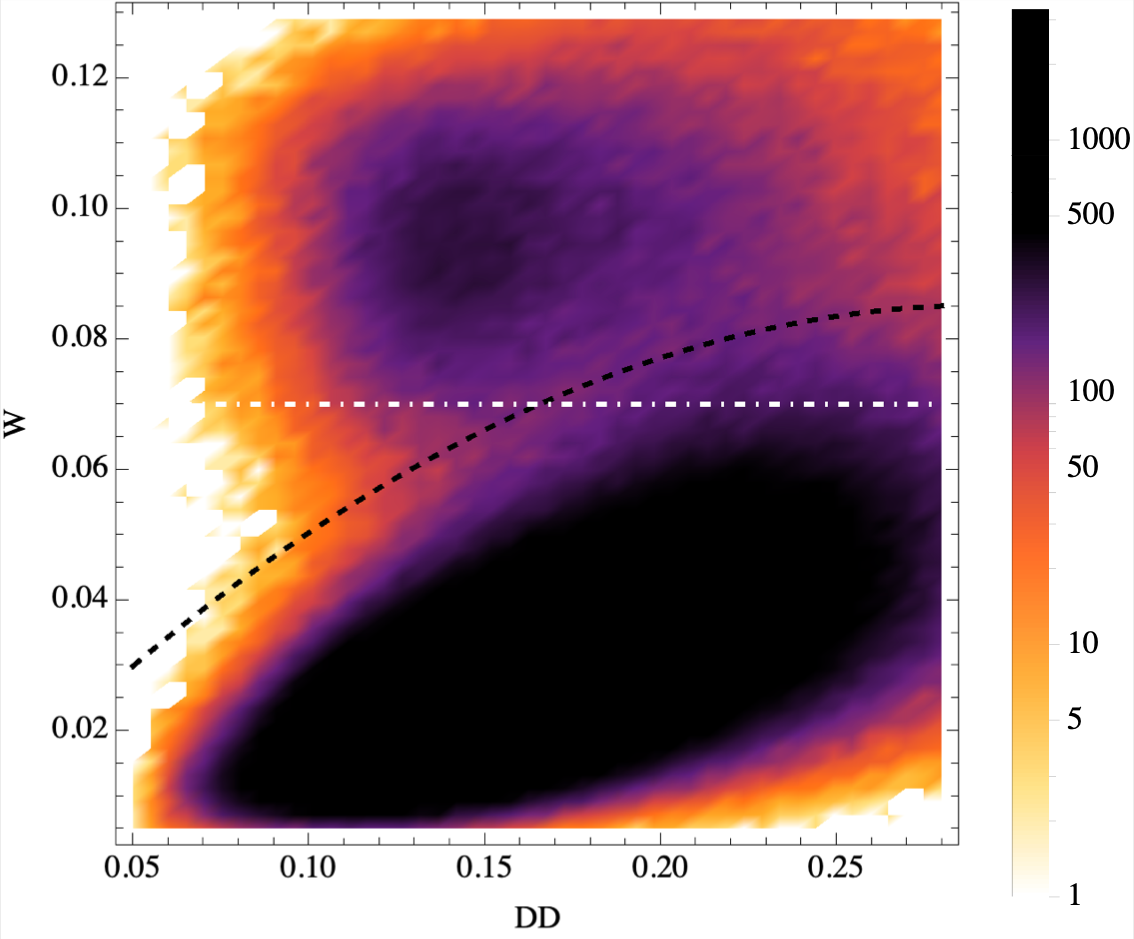}
\end{minipage}
\begin{minipage}{0.04\textwidth}
\centering
\rotatebox[origin=t]{90}{\hspace*{2.5cm} Density \hspace*{2cm} \rotatebox{-90}{{\bf (a)}}}
\end{minipage}
\begin{minipage}{0.435\textwidth}
\centering
\includegraphics[width=\textwidth]{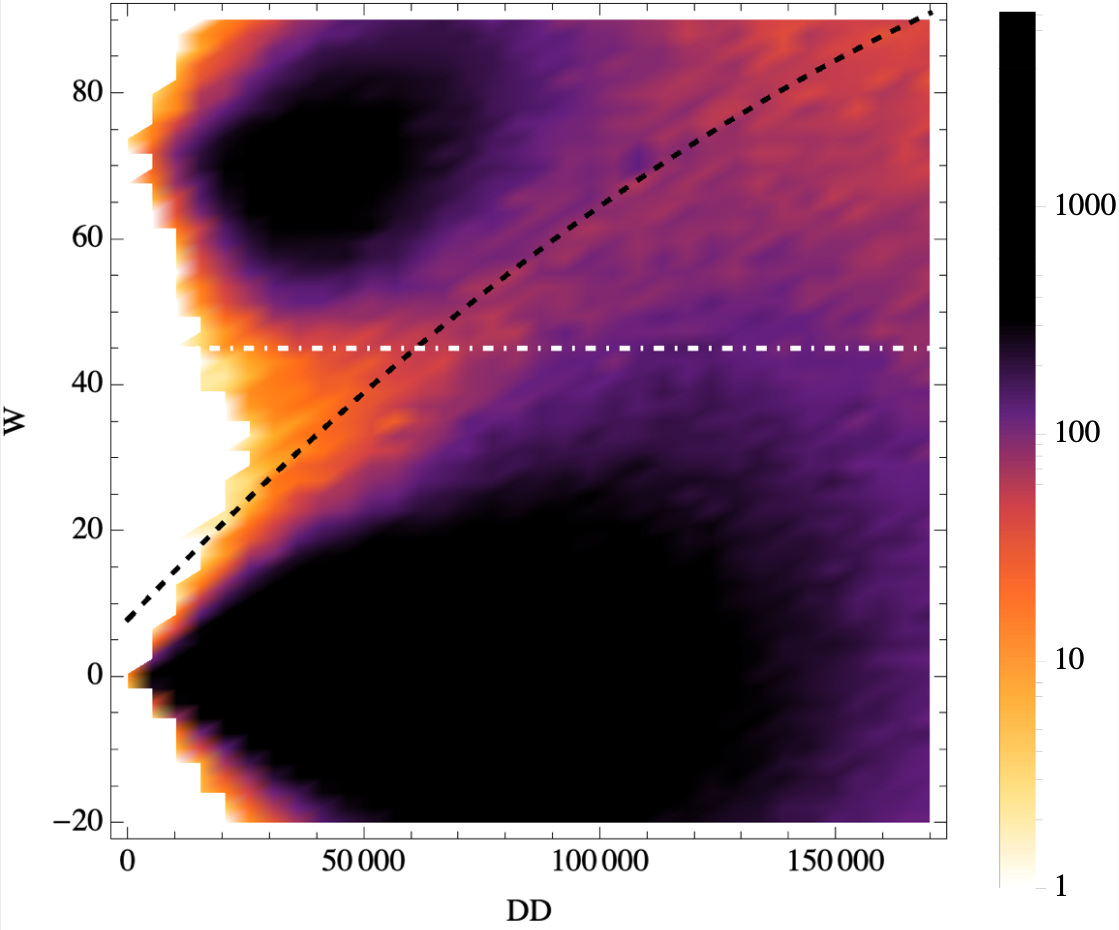}
\end{minipage}
\begin{minipage}{0.04\textwidth}
\centering
\rotatebox{90}{\hspace*{2.5cm} Density \hspace*{2cm} \rotatebox{-90}{{\bf (b)}}}
\end{minipage}
\caption{{\it Mountain-pass} threshold (black dashed line) and a possible choice for the na\"ive threshold (white dotted-dashed line) on top of 
the density histogram for the inferred coupling strengths as a function of the DD for both inference methods $G_1$ (a) and $G_2$ (b). Colour code expresses the density in the logarithmic scale.}
\label{figDetCSdensityCut}
\end{figure}

For the second thresholding, consider Fig.~\ref{fig:scatters}b-d.
In the figure, the na\"ive threshold corresponds to a horizontal separation line. 
We suggest to make use of the extra dimension gained with the new DD measure and consider a separation line that bends and therefore possibly separates true links from non-links more efficiently, i.e. with less false conclusions.
To this aim, we first compute the histogram of the inferred coupling strengths as a function of the DD, see Fig.~\ref{figDetCSdensityCut}.
Then, we calculate the curve that follows the local density minimum between the two bulges of the histogram (black dashed line in Fig.~\ref{figDetCSdensityCut}).
This curve is then used as the new threshold and we refer to it as the {\it mountain-pass} threshold.
The corresponding result of the mountain-pass threshold in terms of false conclusion is illustrated in Fig.~\ref{fig:roc} with a blue triangular marker.
For both $G_1$ and $G_2$, this choice of the threshold results in a better reconstruction of the true links than both the SPL-relative and the na\"ive threshold.

\section{Conclusion}

In this paper, the influence of local network characteristics on the probability of false conclusions about the links inferred from typical
data analysis methods, has been examined.

We considered binary directed networks of coupled oscillators and assumed a setup where only individual nodes can be observed.
Namely, connectivity can not be measured directly, but instead can only be estimated from dynamical observations of individual oscillators. 
The particular methods of connectivity inference adopted in this manuscript take signals of individual nodes and yield a real-valued connectivity matrix representing link weights.
In order to obtain binary connectivity from weighted connections, one would typically threshold link weights to determine their presence. 
A portion of links is almost always misidentified. 
In this paper we investigate the relationship between these false conclusions and local network characteristics. 
In particular we look into two network characteristics: the shortest path length and the detour degree. 
By performing a statistical analysis on simulations where the ground truth is known, we found that these local characteristics can provide additional information regarding the probability of false conclusions. 
The knowledge of the dependency of the inferred link weights and these characteristics allows the links to be represented in a higher dimensional space, where more advanced thresholding techniques can be used. 
Two novel thresholding techniques are proposed as examples, both decreasing the proportion of false conclusions for the tested conditions, see Sec.~\ref{secLast}. 

Additionally, we demonstrated that such \textit{a posteriory} calculated
local network characteristics can provide good estimators of confidence in obtained links, see ROC curves, Fig.~\ref{fig:roc}. 
These results can be applied to real experimental settings, where the underlying true network is not known {\it a priori}. 
As such, these multidimensional thresholding techniques show potential for use in a variety of further investigation. 

In future studies, different reconstruction methods should be considered to check whether the common rules found in this manuscript apply to a wider range of cases. 
Furthermore, deliberating knowledge-based criteria for determining how effective a particular local characteristic serves for such purposes, could lead to conception of optimized network characteristics.

\vspace*{1cm}

\begin{acknowledgements}

This project has received funding from the European Union's Horizon 2020 research and innovation programme under the Marie Sklodowska-Curie grant agreement No 642563. A.P. thanks Russian Science Foundation (Grant Number 17-12-01534).
The authors declare no competing financial interests.
\end{acknowledgements}

\bibliographystyle{apsrev4-1}
\bibliography{biblio.bib}

\end{document}